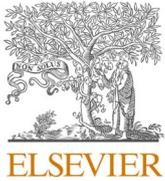
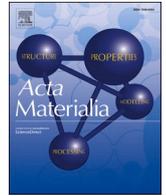

Full length article

# Emerging hierarchical dislocation structures: Insights from scanning electron microscopy-electron backscatter diffraction in situ tensile testing and multifractal analysis

Mikhail Lebyodkin [a,*] , Maxim Gussev [b], Jamieson Brechtl [b], Tatiana Lebedkina [a]

[a] *Laboratoire d'Etude des Microstructures et de Mécanique des Matériaux (LEM3), Université de Lorraine, CNRS, Arts et Métiers Institute of Technology, F-57000 Metz, France*
[b] *Oak Ridge National Laboratory, P.O. Box 2008, Oak Ridge, TN 37831-6136, USA*

ABSTRACT

Understanding the evolution of dislocation structures during plastic deformation is critical for predicting the mechanical performance of metallic materials. In this work, we applied in situ scanning electron microscopy/ electron backscatter diffraction tensile testing combined with multifractal (MF) analysis to assess deformation-induced dislocation structure evolution in solution-annealed 304 L stainless steel, both in its as-received and neutron-irradiated states (5.4 displacements per atom). The analysis of kernel average misorientation patterns revealed the formation of hierarchical dislocation arrangements that exhibit clear MF scaling behavior. Despite pronounced visual differences between nonirradiated and irradiated specimens—most notably, the appearance of dislocation channels after irradiation—the singularity spectra suggest that both conditions give rise to similar underlying hierarchical structures. MF analysis provides a quantitative measure of the spatial complexity and self-organization of dislocation patterns, highlighting the accelerated emergence and evolution of the dislocation structures in irradiated polycrystalline materials, as well as the limitation of their spatial extent. The findings indicate that irradiation not only modifies microstructure but also alters correlation-driven dislocation organization. More generally, they demonstrate that MF analysis is a powerful tool for probing mesoscale deformation mechanisms.

## 1. Introduction

The development of dislocation theory was a groundbreaking advancement that enabled rapid progress in the understanding of plasticity, strain hardening, fracture, and related phenomena. The direct observation of dislocations—particularly through transmission electron microscopy (TEM)—provided insights at scales ranging from a few nanometers to several micrometers. Techniques such as X-ray and neutron diffraction further extended this capability, enabling the analysis of dislocation structures at scales of several millimeters and beyond.

However, the mesoscale that encompasses individual grains or small clusters of interacting grains remained relatively underexplored until the advent of tools such as electron backscatter diffraction (EBSD) and X-ray microdiffraction. These techniques, often described as forms of "dislocation microscopy," have proven highly informative. Notably, EBSD can estimate dislocation densities across length scales from microns to millimeters [1–5]. Yet once such maps are obtained, a critical question arises: How should mesoscale dislocation structures be described and interpreted?

### 1.1. EBSD approach to microstructure evolution

#### 1.1.1. Dislocation structure assessment via SEM-EBSD in situ testing

Conventional EBSD covers scales ranging from about 100 nm to several millimeters or more, depending on the SEM field of view [6]. EBSD and its advanced variant, high-angular-resolution EBSD [1,2], quantify crystal misorientations using multiple parameters, such as kernel average misorientation (KAM) [3] or modified crystal deformation [4]. KAM correlates with the density of geometrically necessary dislocations (GNDs) [7–10], allowing for the mapping of dislocation fields over different length scales. EBSD-retrieved misorientation maps thus provide a quantitative tool to characterize the complex evolution of

M. Lebyodkin et al.Acta Materialia 309 (2026) 122138

dislocation distributions and the effects of different processes, such as irradiation. Because of its ability to reveal grain and subgrain structures in detail, EBSD has been widely employed for studying deformation mechanisms and strain-induced processes [10–12] to derive information on heterogeneous dislocation structures. This capability led to the next logical step: coupling EBSD with mechanical testing. Such a combination allows for in-depth assessment of deformation processes at various length scales, covering a wide range of stress and strain levels [13].

Several limitations of EBSD analysis are also worth mentioning in this context. For instance, EBSD is fundamentally confined to the specimen surface unless destructive methods are employed. In addition, although methods have been proposed to estimate the total dislocation density (see, for instance, [14]), only the GND density can currently be measured reliably, whereas statistically stored dislocations remain difficult to assess. Finally, in situ testing requires interruption of the deformation process for data acquisition, leading to minor relaxation effects.

*1.1.2. Microstructural evolution in irradiated metallic materials*

As high-energy particles (e.g., neutrons in a nuclear power plant) irradiate metallic structural materials, atomic-scale changes first develop in the form of displaced atoms and vacancies—a process widely known as Frenkel pair formation [15]—which alters the arrangement of atoms. From these primary defects, a more comprehensive evolution unfolds over time, resulting in complex defect structures such as dislocation loops, voids, and radiation-enhanced or radiation-induced phases [16]. These defects considerably alter the environment in which dislocation activity develops during plastic deformation. Consequently, interactions between radiation-induced defects and mobile dislocations strongly affect mechanical performance, notably strength, ductility, and deformation behavior [17].

In addition to the formation of crystal lattice defects, metallic materials in radiation environments experience the effect of aggressive conditions (e.g., high-temperature, high-pressure water or molten salt [18]) combined with mechanical and thermal fields [19]. These conditions are aggravated by phenomena such as element transmutation [17], radiation-enhanced diffusion and segregation, and trapped gas accumulation. For a deeper understanding of these and other fundamental aspects of radiation effects on materials, the reader is referred to [15,18,19] and references therein.

The present work considers a practically important case of 304 L stainless steel irradiated at temperatures relevant to light-water reactors (LWRs) (~320 °C). In this context, a key aspect is the comprehensive interaction between the above-described defects in metallic materials and dislocations. For instance, whereas the direct consequence of radiation-induced defects is to impede dislocation motion, thereby increasing strength, dislocation–defect interaction may also lead to specific phenomena such as defect annihilation, resulting in the formation of defect-free channels. Consequently, the deformation mode switches from the formation of multiple fine slip lines in non-irradiated steel to the appearance of a few relatively coarse defect-free channels. The basics of this phenomenon are discussed in [15,18]; for EBSD-related aspects, refer to [6].

Whereas EBSD is an appropriate tool for analyzing microstructure evolution at the grain level or across a group of grains [6,13], the available literature provides either qualitative or semi-quantitative methods that rely on parameters like KAM. Although more quantitative frameworks such as crystal plasticity have been proposed [20], there remains a need for an approach capable of quantifying dislocation interactions and their evolution at the mesoscale (e.g., across multiple interacting grains) or even at the scale of an entire specimen.

*1.2. Multifractal analysis: insight into correlated patterns*

*1.2.1. Fractals and multifractals in nature*

Dislocation structures exhibit complex spatial morphologies that evolve over time. This emerging complexity is difficult to handle using traditional approaches that are based, for instance, on dislocation density alone. To address this challenge, the present work employed the framework of fractals and multifractals (MFs). Fractals are often observed in nature and attract attention due to the imaginative patterns that are caused by their geometries possessing the property of self-similarity (or scale invariance) [21]. This property is defined as a phenomenon where a magnified part of an object (its "fraction") bears a resemblance to the object as a whole [22,23].

Examples of fractal objects in materials science include dendritic growth, martensite, fracture surfaces, and many other complex structures [24]. Scale invariance can be illustrated using a construction on a linear segment (Fig. 1(a)), which is the well-known Cantor "dust" [25]. Its self-similarity arises from an iterative procedure in which the middle third of each segment is removed at every construction step. Covering the segment with a grid then shows that this set obeys the relationship $N \sim 1/\delta l^f$ ($f = 0.631$), where $\delta_l$ is the grid box size and $N$ is the number of nonempty boxes [26]. By contrast, applying the same measurement to the original segment yields the familiar relationship $N \sim 1/\delta l^D$, with the exponent given by the corresponding topological dimension, $D = 1$. The quantity $f$ was termed the "fractal dimension", highlighting the inequality $f < D$. Since we are interested in the analysis of 2D images, Fig. 1(b) presents an example of a fractal structure constructed on a plane, namely the so-called Sierpinski gasket ($f = 1.585$), which is generated by the repeated removal of triangular subsets [27]. Such geometrical constructions exhibit remarkable similarity to real-world objects. For example, Fig. 1(c) shows a fractal structure formed in 2D space by ice crystals grown in supercooled water [28].

Despite the complex visual geometry of these structures, they can be described by a single fractal dimension value. In general, natural objects lack such uniformity, which raises an important issue: although a single fractal dimension, as a global characteristic, can detect clustering, it cannot account for local variations in structural organization [26]. In many cases, multiple fractal dimensions are required to properly describe the underlying self-similar geometries [29–31]. Such structures may lack visually recognizable patterns but nevertheless exhibit scale invariance in a statistical sense.

This complexity is often due to a nonuniform distribution of a physical quantity over the geometrical support, for example local magnetic moments [32] or the instantaneous intensity of acoustic emission (AE) during plastic deformation [33,34]. To describe this distribution, the corresponding physical quantity, $\mu$, is used to define a probabilistic measure. In the above example of Cantor dust, $\mu$ can be implemented by replacing the geometrical segments with rods carrying

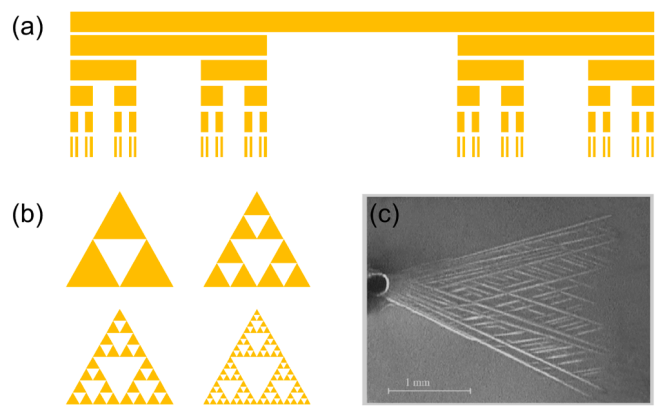

**Fig. 1.** Examples of fractal structures. (a) Five iterations (progresses downward) of the fractal Cantor dust created by iteratively removing the middle third of each segment in the current construction [25,29], (b) four iterations of the Sierpinski gasket created by iteratively removing a center triangle from the current construction, (c) ice crystals grown in supercooled water (reproduced with permission from [28]).

a mass.

The generalization from the monofractal to the MF can be demonstrated using a simple example known as the binomial cascade, which is illustrated in Fig. 2(a) [35,36]. The construction starts from a rod with a unit mass. The iterations consist of partitioning a given segment into two parts and altering their masses in accordance with fixed rules. For the present example, the masses of the first and second parts are reduced by values of $\mu_i$ ($\mu_i < 1$) and $1 - \mu_i$, respectively. Fig. 2(b) and (c) show, respectively, the absolute value of the time-dependent stress exhibited by a high-entropy alloy (HEA) under tensile loading and a random MF based on intermittent turbulence, as introduced by Mandelbrot [37,38]. The similarity between the binomial cascade and these two examples demonstrates that simple rules may underlie very complex behavior, thereby opening avenues for quantitative characterization. Moreover, as described in Sec. 1.5, MF complexity may arise from nontrivial scaling properties of the measure itself, even when it is distributed over a nonfractal geometrical support. This situation is rather the rule than the exception for natural objects, owing to experimental noise, unless physically based denoising can be applied—an approach that is not trivial for complex signals or images with unknown structure.

### 1.2.2. Fractality in plasticity and fracture problems

Although fractal analysis has not yet been widely applied to plasticity problems, some promising examples have been published. These include 1D time series obtained from deformation curves (see Fig. 2(b)) or AE signals [33,34,39–41], 2D patterns represented by dislocation structures analyzed via TEM [42–44], deformation-induced surface roughness [45–47], slip band morphology [48], and fracture surfaces [49–51]. As far as plastic deformation patterns are concerned, various types of fractal dimensions have been explored (e.g., the box-counting dimension characterizing the scaling of spatial coverage in the image, the roughness exponent reflecting the scaling of profile height differences, and the correlation dimension describing pair correlations [24, 26]). Despite these differences, the analyses led to qualitatively consistent conclusions: It is generally agreed that dislocation structures and the associated strain heterogeneity evolve into fractal patterns after some deformation, with their fractal dimension increasing during deformation and eventually tending toward saturation [45]. However, the latter conjecture has not yet been established with certainty.

Applications of MF analysis to plasticity and fracture are rare. However, MF analysis has demonstrated its ability to uncover fundamental physics in 1D cases, such as serrated deformation curves resulting from unstable plastic flow and AE signals accompanying deformation [33,52–56]. Moreover, MF analysis has revealed that small stress fluctuations, which are usually regarded as noise, may exhibit a nonrandom structure, suggesting that they arise from correlated physical processes [57]. More recently, the MF analysis of small stress fluctuations has identified a potential indicator of either the imminence of a macroscopic plastic instability [52] or transitions between different types of such instabilities [54].

Despite the body of work regarding MF analysis, it still has not yet been applied to 2D patterns. The lack of these results is likely due to the complexity involved in interpreting the results. Moreover, even the more conventional fractal formalism has not been used to analyze EBSD results.

In the present work, MF analysis is applied to uncover potential correlations in the KAM maps that emerge during plastic deformation of 304 L stainless steel, a material for which mechanical properties are well known. Access to the evolution of dislocation structures is achieved through the implementation of in situ EBSD measurements. Furthermore, both experimental and theoretical analyses are performed not only on the as-received material but also after neutron irradiation in nuclear reactors. Beyond addressing practical questions related to the use of irradiation to examine mechanical properties, this approach enables a comparison of dislocation self-organization at the mesoscale in qualitatively different microstructural states of the same material, thereby providing insight into the fundamental aspects of collective dislocation behavior.

## 2. Material, experiment, and data analysis

### 2.1. Steel element composition, specimen geometry, and in situ mechanical testing

Solution-annealed 304 L stainless steel was used in the present work (see Table 1 for the elemental composition). The reasoning behind selecting this material was three-fold. First, 304 L steel is a practically important material and a typical representative of the 300 series steel class. Second, after annealing, this steel, compared with materials such as hardened 718 alloy or aged 6061 alloy, consists of a "pure" austenite matrix that is mostly free of secondary phases and precipitates that can interact with dislocations. Third, radiation effects in this material are well studied for typical LWR conditions; irradiated 304 L stainless steel has a matrix populated by small dislocation loops, point defect clusters, and some precipitates that can interact with dislocations.

A set of tensile samples designated for stress corrosion cracking experiments [58] was irradiated in the BOR-60 liquid sodium fast reactor to various damage levels at approximately 330 °C for a damage rate of $9.4 \times 10^{-7}$ displacements per atom (dpa)·s$^{-1}$ [59]. The geometry and dimensions of these samples are shown in Fig. 3(a). A sample irradiated to 5.4 dpa was employed in the present study. Ultra-miniature irradiated tensile specimens [60] for in situ mechanical testing, with gauge dimensions of 1.5 mm (length) by 0.6 mm (width) by 0.6 mm (thickness), were fabricated from thin slices cut from the larger tensile sample heads (see Fig. 3(b)). A reference specimen of identical geometry was produced from an archive (non-irradiated) heat of the same material.

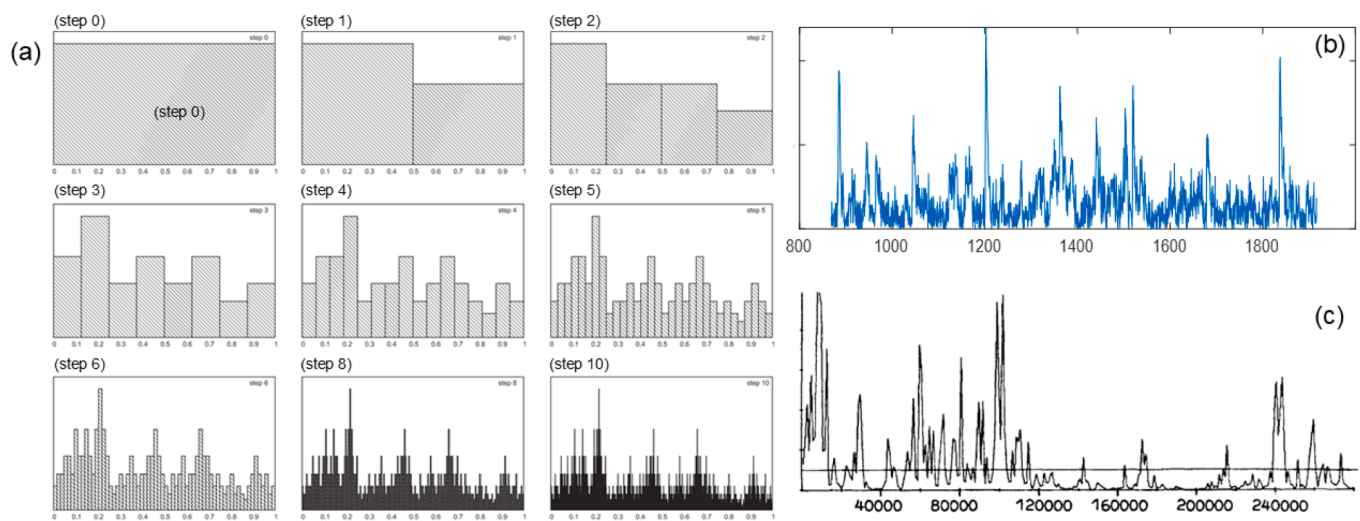

**Fig. 2.** Generalization to multifractals. (a) Graph depicting the binomial cascade produced over 10 iterations, (b) the absolute value of stress fluctuation data for a HEA subjected to tension, and (c) a random MF structure based on intermittant turbulence (reproduced from [36,38] with permission).

**Table 1**
Element composition (weight percent (wt. %), Fe = bal.) of the studied 304 L steel.

| Element | C | Si | Mn | Ni | Cr | Cu | Co | S | P | B (ppm) | N (ppm) |
|---------|---|----|----|----|----|----|----|---|---|---------|---------|
| wt % | 0.022 | 0.68 | 1.79 | 9.88 | 18.61 | 0.25 | 0.064 | 0.0007 | 0.032 | 9 | 610 |

Custom grip adapters (Fig. 3(c)) were designed specifically to accommodate this tensile specimen geometry. The in situ mechanical tests were performed at room temperature at a nominal strain rate of $10^{-3}$ s$^{-1}$ (displacement rate of 1.5 μm/s) using a miniature Kammrath & Weiss (K&W) 5 kN tensile frame (Fig. 3(d)) installed inside a TESCAN MIRA3 SEM. The SEM was equipped with a high-speed Oxford Symmetry EBSD detector (detector resolution: 1244 × 1024 pixels), and experimental data were collected using Oxford AZtec software. To create smooth, clean, and EBSD-suitable surfaces, the tensile specimens were mechanically ground and polished per standard metallography practices, using a Minimet device, down to 1 μm diamond sandpaper. After that, the specimens were electropolished using Struers A2 solution at 30 V DC for 5–7 s. To preserve surface quality, the specimens were stored in plastic membrane boxes prior to the test. During testing, the SEM was operated at an accelerating voltage of 20 kV and a beam current of approximately 4 nA. The working distance was about 25 mm, which was constrained by the in-chamber geometry (Fig. 3(e)).

An in situ mechanical test was performed in a step-by-step fashion, with deformation interruptions to collect EBSD data and SEM images at each step. The zero step, [S00], provided reference datasets; the deformation continued until the EBSD dataset experienced significant degradation caused by surface roughness and an increase in the dislocation density. Per safety requirements, no fracture inside the SEM was allowed for the irradiated specimens, so the strain level was limited to minimize fracture risks. Several regions of interest (ROIs) were selected, scanned prior to the test, and subsequently tracked and rescanned throughout the experiment at different strain and stress levels. Details and methodological aspects of a typical in situ test are provided in [61].

After the test, the experimental datasets were processed using EDAX OIM 8.0 commercial software and specialized software scripts. Attention was given to ensuring correct data transfer and axis orientations. Because the plastic strain level varies along the gauge, the retained plastic strain value calculated from the tensile curve is not a representative parameter, especially for irradiated steel prone to deformation localization [61]. Here, local plastic strains in the ROIs were measured by tracking fiducial marks on the specimen surface and triple junction points in the EBSD maps, similarly to the approaches used in [6,61].

Fig. 4 illustrates the typical microstructure of the studied 304 L steel, which is characterized by a predominantly annealed austenitic matrix with randomly oriented, well-formed grains with an average grain size of about 25 μm. No significant texture was observed, and the fraction of annealing twin boundaries (special Σ3 boundaries) was approximately 50 %. A small volume fraction (<0.2 %) of retained ferrite was also present; most ferrite grains were small and rounded, typically measuring 3–5 µm or smaller. Overall, these features are characteristic of annealed or hot-rolled austenitic 300 series steels.

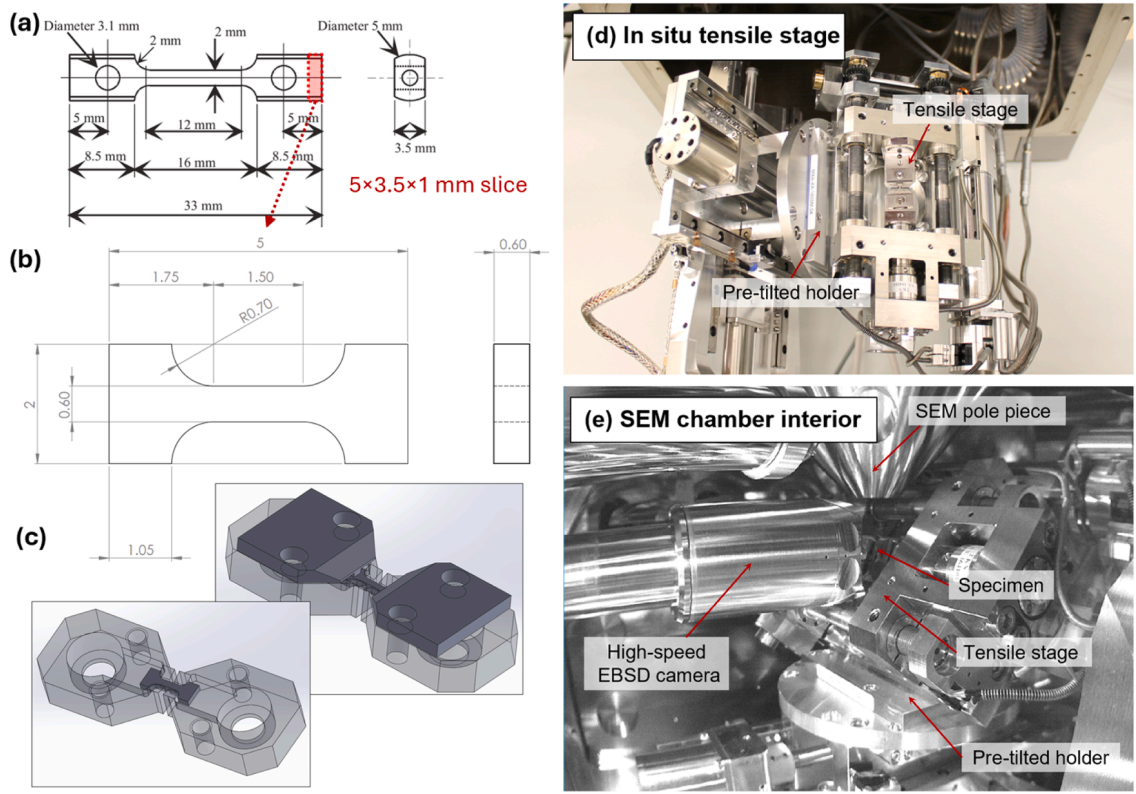

**Fig. 3.** In situ testing. (a) Irradiated material source (round tensile bar) and slice orientation, (b) ultra-miniature tensile specimen geometry (dimensions are in millimeters), (c) grips with tensile specimens with and without holders (CAD view), (d) K&W 5 kN tensile stage installed in the TESCAN MIRA3 SEM, and (e) SEM chamber interior with the stage in the working position and EBSD camera inserted.

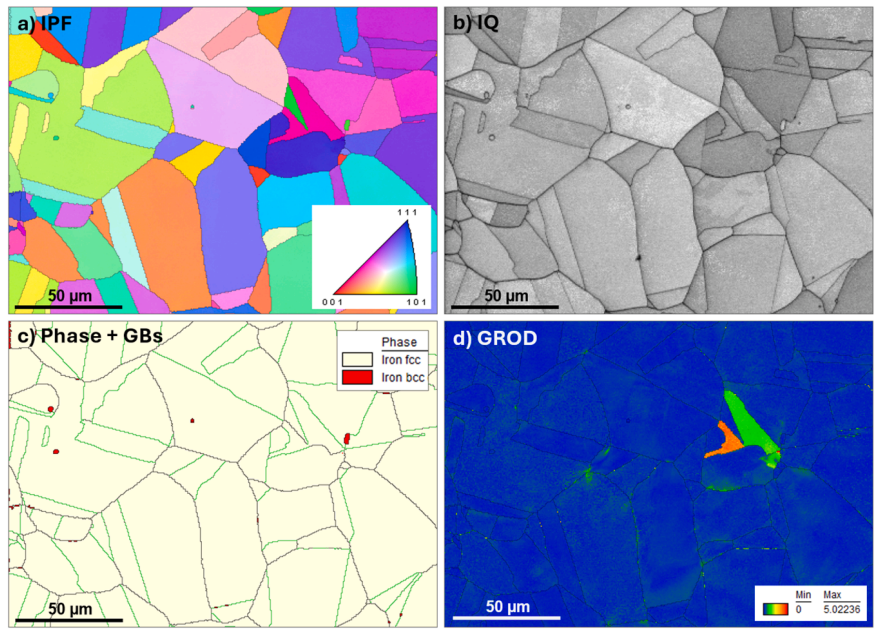

**Fig. 4.** Representative microstructure of the investigated 304 L steel prior to deformation: (a) EBSD inverse pole figure (IPF), (b) image quality (IQ), (c) phase, and (d) grain reference orientation deviation (GROD) maps. For this characterization an EBSD step size of 500 nm was used. The phase map is overlapped with a color-coded grain boundary map (random high-angle boundaries [RHABs] are black; twin boundaries [$\sum 3$] are green, and the fraction of twin boundaries is ~50.1 %). The IPF map is colored in the tensile direction (horizontal); the IPF color key is the same for all IPF maps in this publication.

### 2.2. Multifractal procedure

Conventional KAM map representation is a color-coded plot, intended to highlight variations in local misorientation. Being informative, and visually rich, color coding tends to smooth sharp changes and local gradients. Such gradients may carry additional information on strain-induced phenomena. The discussion below provides the basic formulae employed to analyze the MF features of these KAM maps that are represented as 2D digital images (matrices). To test the scaling properties of the image, the analyzed KAM map is covered by a 2D grid containing boxes with side length $\delta l$ and a local probabilistic measure $\mu_i(\delta l) = \sum_j |\psi_i| / \sum |\psi|$, where $|\psi_i|$ is the absolute value of the ith KAM data point in the jth box [62]. The term in the denominator corresponds to the sum of all $|\psi|$ contained in the entire KAM image. The self-similarity is then assessed by analyzing the scaling of the families of partition functions corresponding to different values of the order, $q$ [57]:

$$Z_q(\delta l) = \sum_i^N \mu_i^q(\delta l), \ q \neq 1; \ Z_1(\delta l) = \sum_i^N \mu_i(\delta l) ln \mu_i(\delta l), \ q = 1 \quad (1)$$

For a MF pattern these partition functions obey a scaling law in the limit $\delta l \to 0$:

$$Z_q(\delta l) = \delta l^{(q-1)D_q}, ; q \neq 1; \ Z_1(\delta l) = D(1) ln(\delta l), \ q = 1 \quad (2)$$

where $D_q$ is the generalized fractal dimension [31].

Next, the singularity strength, $\alpha$, of the local measure can be defined by employing the scaling of the self-similar measure with regard to the box size, $\delta l$ [26,63]:

$$\mu_i(\delta l) \sim \delta l^\alpha \quad (3)$$

The subsets formed from the boxes corresponding to singularity values in the range ($\alpha$, $\alpha + d\alpha$) can be characterized by counting their number $N(\alpha)$ and calculating the respective fractal dimension, $f(\alpha)$, of the analyzed subset [31]:

$$N(a) \sim \delta l^{-f(a)} \quad (4)$$

These new definitions provide clearer ways to represent the MF spectrum as compared with that defined by $D_q$. In particular, $f(\alpha)$ corresponds to the fractal dimension of the subset of boxes corresponding to the singularity strength in the vicinity of $\alpha$. In other words, it describes the heterogeneous object as interpenetrating fractal subsets [57]. It is useful to note that this description highlights the difficulty of interpreting the MF analysis of a real object, as the subset corresponding to a certain fractal is composed of parts of the object from different sites.

It is also important to point out that the box-counting dimension, equivalent to the form $f_{max} = D(0)$, is close to the topological dimension because the images are treated as is, including the noise. This approach was chosen because it allows arbitrariness related to denoising to be avoided. In the present study, the following direct method was used to calculate $f(\alpha)$. This method is based on scaling relationships for a normalized measure, $\widetilde{\mu}_i(\delta l, q) = \mu_i^q / \sum_j \mu_j^q$ [64]:

$$\Sigma_\alpha(\delta l, q) = \sum_i \widetilde{\mu}_i(\delta l, q) ln \mu_i(\delta l) \sim \alpha(q) ln \delta l$$

$$\Sigma_f(\delta l, q) = \sum_i \widetilde{\mu}_i(\delta l, q) ln \widetilde{\mu}_i(\delta l, q) \sim f(q) ln \delta l \quad (5)$$

It is important to specify that due to the rapid growth of the power law function, variation in $q$ is equivalent to marking specific subsets corresponding to distinct exponents in Eq. (2–4). For example, choosing large positive or large negative $q$ values makes the subsets corresponding to the largest and smallest $\mu_i$ values dominate in $Z_q$, respectively. Varying $q$ between two limits allows for scanning various subsets, much like how, when using a microscope, the observer continuously changes the distance between the objective and the object, bringing finer or coarser details into contrast. This property is known as a 'mathematical microscope'. It follows that, when applied to real noisy objects, the MF analysis has a limited ability to interpret results for $q < 0$ because the corresponding subsets with the smallest measures are the most difficult to evaluate in experiments. For this reason, the results presented in the following sections focus only on the $q > 0$ branches.

Some values of $D$, $f$, and $\alpha$ allow a direct physical interpretation [63]. Among these, $f_{max} = D(0)$ corresponds to the fractal dimension of the support on which the measure $\mu$, as defined above, is distributed over the KAM maps. The case $q = 1$ corresponds to the information dimension, for which $D(1) = \alpha(1) = f(1)$, and characterizes how rapidly the underlying information increases with increasing resolution. The correlation dimension $D(2)$ reflects pairwise correlations between local misorientations. It might be noted for clarity that, since the analysis is applied to as-measured KAM maps without removal of experimental noise, $f_{max}$ is close to the topological dimension $D = 2$, because taking $q = 0$ in Eq. (1) corresponds to counting all nonempty grid boxes.

Essentially, the span of $\alpha$ between the summit and the edges of the $f(\alpha)$ spectrum reflects the heterogeneity of the analyzed structure. As the formation of a hierarchical structure during deformation lies at the core of the present study, it is the evolution of this spectrum as a whole that is primarily examined in what follows. Another important quantity is the upper scaling limit, which makes it possible to relate the size of the regions where hierarchical structures form to specific regions observed in EBSD patterns, such as those shown in Fig. 4.

## 3. Experimental results and analysis

### 3.1. Mechanical behavior and strain induced changes in the microstructure of reference and irradiated specimens

Fig. 5 shows experimental tensile curves with strain steps performed to collect EBSD data and record SEM images. Strain interruptions are visible as force drops (small load decrements due to minor relaxation effects in the material). The yield stress for the irradiated specimen is significantly higher, reaching approximately 900 MPa, reflecting radiation hardening [16]. Both the irradiated and reference specimens did not reach the necking stage where macroscale strain localization develops. Note that the tensile curve for the irradiated specimen shows a force drop after yielding and a barely noticeable increase in the stress with deformation after the force drop. Force drops and strain-induced softening are often observed in irradiated austenitic steels [65], sometimes forming a long plateau in the tensile curve. In this case, engineering tensile curves are not the best tool to study hardening behavior, and "true stress – true strain" curves are usually employed [66]. This aspect is not discussed in the present work.

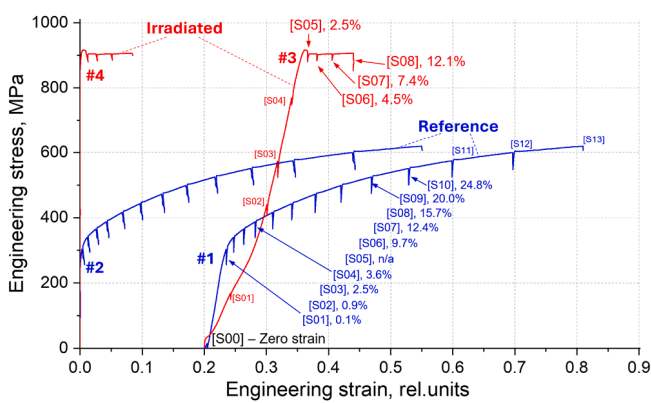

**Fig. 5.** Experimental tensile curves for the non-irradiated (#1, #2; blue curves) and irradiated (5.4 dpa, 330 °C) (#3, #4; red curves) 304 L steel specimens. #1, #3: Engineering strain vs. engineering stress (note that the curves are shifted to the left to aid the reader). #2, #4: Engineering plastic strain vs. engineering stress (elastic compliance removed). Both step number (e.g., [S00] and so on) and local strain inside the ROI are shown. Because the datasets at steps 1–4 for the irradiated specimen were intended for studying micromechanical effects, they are not discussed in the present work. Maps for steps 11–13 of the reference specimens are missing a significant fraction of data points and therefore are not suitable for MF analysis; the step 5 dataset was compromised because of hardware failure.

Fig. 6 shows the experimental EBSD maps recorded for the same ROI during the in-situ experiment. The color-coded IPF map represents grain morphology and orientation, whereas the KAM map reflects the distribution of GNDs. The IPF map recorded at zero strain level, prior to plastic deformation, shows grains that are mostly free of in-grain misorientation gradients, (see Fig. 6(a)). The corresponding KAM map, shown in Fig. 6(b), demonstrates a noise-like pattern that reflects the overlapping of EBSD measurement noise and randomly distributed dislocations. At small plastic strains, as shown in Fig. 6(c)–(d), minor color variations become evident in the IPF map, reflecting lattice rotation processes and formation of dislocation arrays. Furthermore, the KAM map shows weak changes at this strain level, and the overall picture may be interpreted as the activity of multiple dislocation sources and development of fine slip lines. Additionally, the spacing between slip lines is smaller or comparable to the EBSD step size, so single slip lines cannot be resolved. Finally, the dislocation activity inside slip lines manifests itself in lattice rotation.

At moderate strains, as shown in 6(e)–(f), color variations inside the grains become complex. Such behavior reflects dislocation movement and evolution within multiple slip systems, as well as complex dislocation interaction and accumulation. The KAM map shows specific "bands" along grain boundaries; the bands are the result of dislocation accumulation near obstacles (grain boundaries). Color variations in the KAM maps are also observable near triple junction points.

At large strains, which correspond to Fig. 6(g)–(h), color variations in the IPF map become stronger. This trend reflects the increasing in-grain misorientation and, for some grains, the initial stages of grain fragmentation. Furthermore, the KAM map reveals growing dislocation density in which complex structures arise at grain boundaries and inside grains, reflecting slipping activity within several active slip planes.

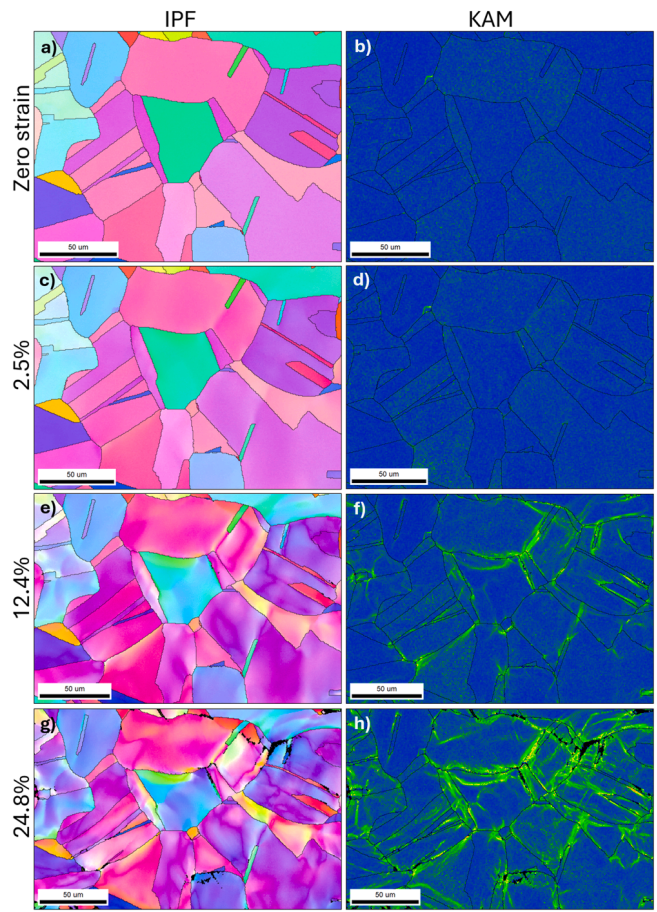

**Fig. 6.** EBSD maps showing microstructure evolution in reference (non-irradiated) specimen at different strain levels (0–24.8 %).

Overall, the IPF and KAM maps recorded at different strain levels for the same area suggest very complex deformation mechanisms. Although some aspects (e.g., lattice rotation and dislocation arrays along grain boundaries) are easy to interpret, this interpretation is mostly qualitative. Even if the dislocation density is calculated from the KAM parameters, this value provides limited information on the spatial distribution and interactions of dislocations.

Fig. 7 shows the EBSD dataset for the 5.4 dpa irradiated specimen. The reference maps presented Fig. 7(a)–(b) appear to be very similar to those of the non-irradiated specimen that are presented in Fig. 6(a)–(b). Conventional EBSD is not sensitive to the formation of "black dots", dislocation loops, or other radiation-induced defects with sizes on the order of 1–10 nm. The insensitivity of the EBSD is due to its beam interaction area having dimensions of ~100–120 nm, (at 20 kV for a 70°-tilted specimen), which as a rule make smaller features, such as those described above, practically invisible.

A small plastic strain of 2.5 % (see Fig. 7(c)–(d)) produces misorientation "spots" (local variations in orientation) in the IPF map and strong localized fluctuations in the KAM map. The overall pattern—a few "hot spots" per grain with strong KAM variations—is in striking contrast with non-irradiated steel at the same strain level (Fig.6(c)-(d)). This observation reflects a transition from multiple slip lines in the reference material to a few coarse dislocation channels in the irradiated one [6]. Deformation localizes inside channels, leading to strong dislocation pile-ups at obstacles (e.g., grain boundaries). Therefore, at some point, backstress from the pile-ups may cease activity within the channel [66]. As deformation progresses (see Fig. 7(e)–(f)), more hot spots form, reflecting the formation of new channels [67]. At a local strain of ~12.1 %, as found in Fig. 7(g)–(h), the irradiated material shows multiple delta- or triangle-like spots of localized deformation located mainly along grain boundaries. Additionally, hot spots form inside grains at channel intersections.

### 3.2. MF analysis of the EBSD datasets

Fig. 8 presents examples of families of partition functions, $Z_q/(q-1)$, as a function of the grid box size $\delta_l$, illustrating the strain-induced emergence of long-range scaling in the KAM patterns. At large $\delta_l$, all dependencies converge to a single straight line with a slope of 2, indicating globally non-fractal behavior in 2D space. Some undulations in this scale range may arise, for example, from $\delta_l$ approaching the EBSD image size, resulting in only a few grid boxes fitting within the entire image [31], or from variations in the KAM intensity between different grains (see Fig. 6). Essentially, the dependencies deviate from this trivial trend and become $q$-dependent as $\delta_l$ decreases. The curves shown in Fig. 8(a) for zero strain can be approximated by straight lines with varying slopes over a very short $\delta_l$ interval. These slopes, which range from 1 to about 5 pixels, are illustrated by the dashed line along the upper curve. It seems reasonable to suggest that the KAM distribution in the undeformed sample is uncorrelated, except perhaps at very short distances.

The plots for the initial deformation steps are not shown here because they were qualitatively similar to Fig. 8(a), consistent with the similarity of the corresponding KAM images. However, in contrast to the visual comparison of the KAM patterns, the quantitative analysis of partition functions revealed gradual changes, including the emergence of a second scaling interval at larger $\delta_l$. The presence of two distinct families of straight dependencies is clearly visible in Fig. 8(b), where two dashed lines indicate the scaling intervals identified for $\varepsilon = 3.6$ %. With further deformation, a unique scaling behavior emerged over a broad $\delta_l$-interval, as illustrated in Fig. 8(c) for the maximum strain of 24.8 %. Overall scaling was observed for all images from $\varepsilon = 9.7$ % onward.

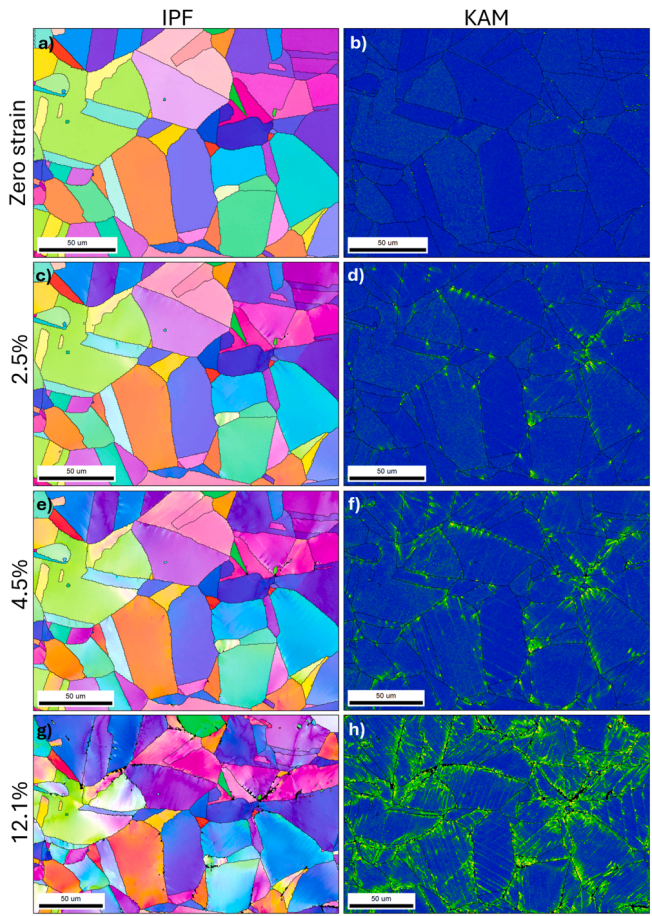

**Fig. 7.** EBSD maps showing the microstructural evolution in the irradiated specimen at different strain levels (0–12.1 %).

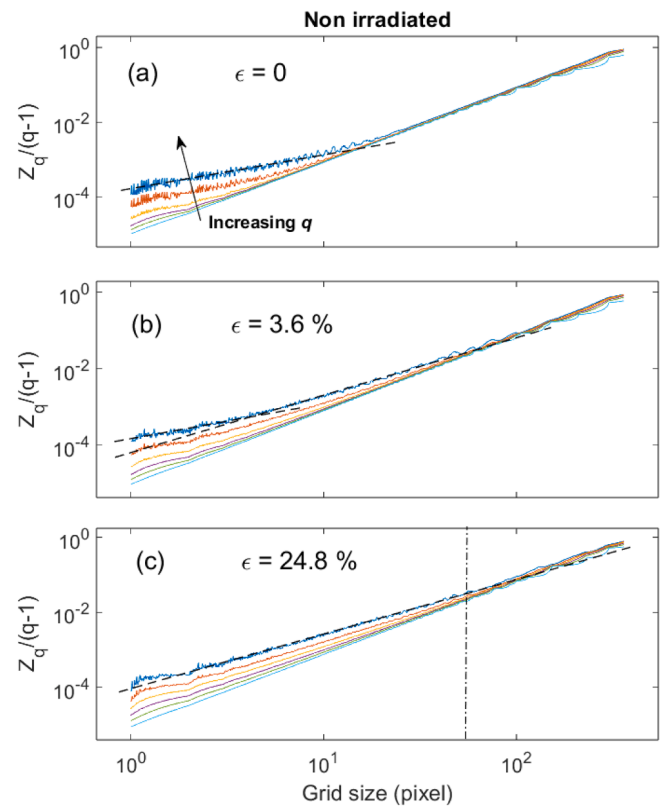

**Fig. 8.** Families of partition functions for three strain levels in the non-irradiated sample (see Figs. 5 and 6). Examples are shown for several q-values (between 0 and 20, increasing in the direction of the arrow). Dashed lines indicate power-law dependencies, and the vertical dash-dotted line marks the inflection point from the corresponding straight lines. Panels (a), (b), and (c) correspond to the strain levels indicated directly in the plots.

The lower scaling limit typically extended to one or two pixels, constrained solely by the experimental resolution. The evolution of the upper scaling limit is discussed in detail below. Notably, this upper bound ranged from approximately 40 to 60 pixels (20 to 30 µm), aligning with the variation in grain size. It can therefore be inferred that deformation gives rise to a hierarchical GND structure with fractal characteristics, displaying similar scaling behavior across different grains. However, the scaling limitation is likely due to variations in the dislocation activity between grains, which disrupt the continuity of the evolving GND structure. It is also noteworthy that the continuous evolution of partition function families from the undeformed state to the maximum strain supports the earlier remark concerning possible scaling at short distances. This suggests that the tendency toward scaling behavior over a short $\delta_l$ interval at zero strain is unlikely to be an artifact—such as $\delta_l$ approaching the size of a single pixel (see [33])—but instead reflects short-range correlations within the initial GND structure of the material.

The emergence of scaling behavior is clearly evident in the singularity spectra, $f(\alpha)$, calculated from the data illustrated in Fig. 8. Fig. 9(a) displays $f(\alpha)$-curves within the statistically significant range of non-negative $q$-values for several strain levels, highlighting meaningful changes in the spectra. Let us first examine the $f(\alpha)$ dependence at zero strain. As described in Section 2, the MF formalism characterizes self-similar objects through continuous $f(\alpha)$-functions confined to the $f \geq 0$ half-plane. Consequently, the left-side wing of this dependence, corresponding to large $q$-values, is unphysical ($f < 0$) and indicates non-fractal behavior. In other words, the subsets where the KAM density is most concentrated—dominating the partition functions at large $q$—are not structured in a way that obeys scaling laws. However, in line with the above remarks on the tendency of partition functions to exhibit scaling behavior for the undeformed sample, the range of $q \leq 3$ displays smooth behavior, consistent with the changes in $f(\alpha)$ observed during subsequent deformation. This trend reinforces the hypothesis of a certain degree of non-randomness in the initial GND distribution. Furthermore, the gradual nature of the subsequent changes is clearly illustrated by the analysis at a strain of 3.6 %, which corresponds to the presence of two distinct scaling intervals (see Fig. 8(b)). Notably, the left-side wing of $f(\alpha)$ becomes smoother, and the right-side wing develops a spectrum with a finite width. It is worth recalling that non-fractal behavior, characterized by a uniform slope of 2 for all $q$-values, results in the collapse of the singularity spectrum to a single point at $\alpha = 2$ and $f = 2$.

In Fig. 9(a), the $f(\alpha)$ dependence at 9.7 % strain—the point at which an overall scaling behavior has emerged—exhibits a well-defined spectrum across the entire $q$-range. Interestingly, a slight imperfection, resembling an angle around $f \approx 0.5$, is still present at this stage. This result suggests that the development of the hierarchical structure remains incomplete. It can be assumed that this irregularity arises from the contrast between highly singular misorientations at grain boundaries and the less singular behavior within the grains. Still, the spectrum continued to evolve during subsequent deformation, becoming smoother and slightly narrower and thus indicating reduced heterogeneity (see the curves for 12.4 % and 24.8 % strain). Overall, the changes seem to tend toward saturation at large strains. This finding suggests that the formation of a global hierarchical GND structure within the grains gradually reached completion.

The strain range between 12.4 % and 24.8 % is of particular interest because it corresponds to well-determined singularity spectra that vary only weakly over a broad strain interval. It is thus instructive to compare the KAM maps corresponding to these limits (Fig. 6(f) and (h)). Despite the small quantitative differences revealed by the MF analysis, the KAM images exhibit pronounced visual changes. By contrast, although the deformation between 9.7 % and 12.4 % strain corresponds to a transition from an imperfect to a smooth singularity spectrum, the corresponding KAM images remain largely similar; the 9.7 % image is therefore not shown in order to avoid overloading the figure. Taken together, these observations suggest that the apparent evolution of the KAM maps is primarily associated with the accumulation of dislocations near grain boundaries, whereas the MF analysis provides access to the underlying hierarchical dislocation structure within grains that is not readily apparent from visual inspection. This result highlights the need for quantitative approaches in the analysis of EBSD patterns.

Figs. 9(b) and 10 present analogous results for the irradiated sample. The family of partition functions shown in Fig. 10(a) for the undeformed material, along with the singularity spectrum in Fig. 9(b), are qualitatively similar to their counterparts in Fig. 9(a) and 8(a), which display the corresponding data for the non-irradiated sample. Accordingly, the hypotheses regarding possible short-distance correlations in the undeformed state, discussed above for the initial material, remain relevant post irradiation.

Compared with the non-irradiated material, the irradiated material exhibits overall scaling behavior of the partition functions and a regular singularity spectrum as early as 2.5 % strain (see Fig. 9(b) and 10(b)). A unique scaling behavior is also observed at higher strains. Interestingly, despite the visual differences between the GND distributions in the two materials, as discussed in Section 3.1, their singularity spectra exhibit rather similar MF properties. Indeed, once a structure characterized by global scaling is established, its further evolution results in the narrowing of $f(\alpha)$ toward a saturated dependence, with the value of $\alpha_{min}$ (the left edge of the $f(\alpha)$ spectrum) evolving within a similar range in both cases. It is also worth noting that in both materials, the $f(\alpha)$ dependencies observed at relatively early deformation stages exhibit similar imperfections, as seen when comparing the dependencies at 9.7 % strain in the nonirradiated sample and at 2.5 % in the irradiated material.

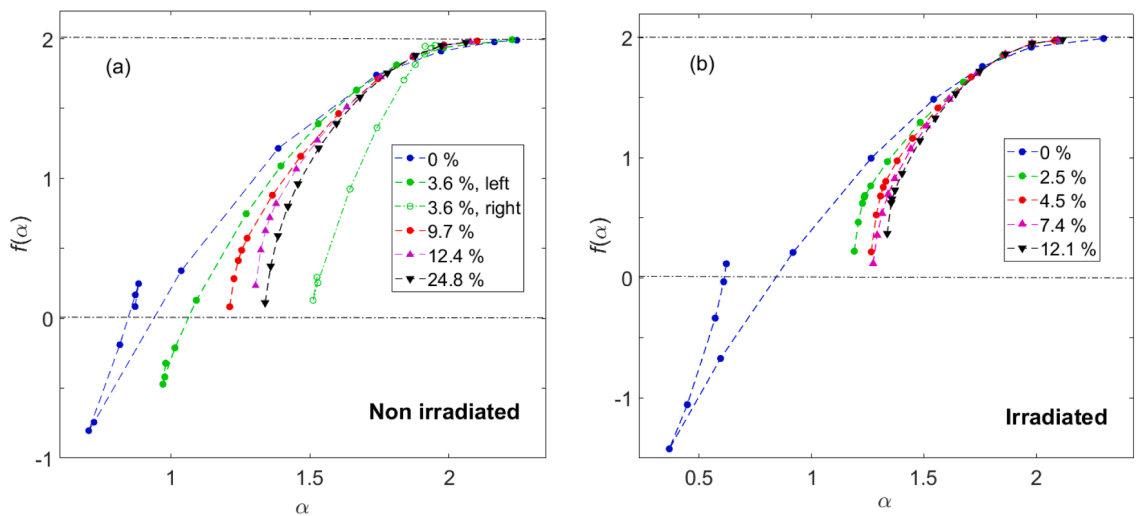

**Fig. 9.** Singularity spectra $f(\alpha)$ for the range $0 \leq q \leq 20$ for different strain values (see legend). (a) Initial sample—some intermediate dependencies are omitted to maintain plot readability. (b) Irradiated (5.4 dpa, 330 °C) sample.

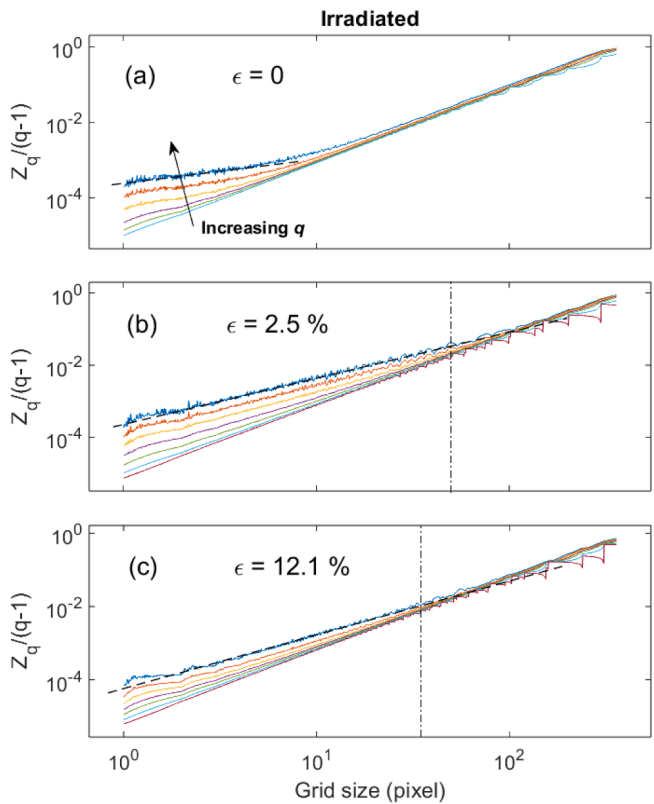

**Fig. 10.** Families of partition functions for three strain levels in the irradiated sample (5.4 dpa, 330 °C; see Figs. 5 and 7). The data are presented following the same conventions as in Fig. 8, including *q*-values, power-law fits (dashed lines), and inflection points (vertical dash-dotted lines). Panels (a), (b), and (c) correspond to strain levels indicated directly in the plots.

The similarity of the MF spectra suggests that the fundamental hierarchical structure induced by plastic deformation is similar in both cases. This conclusion is unexpected, given the obvious visual differences between the KAM patterns in Figs. 6 and 7. However, this apparent contradiction can be explained simply. Indeed, although they correspond to similar scaling indices, the partition functions shown in Figs. 8 and 10 exhibit two key differences. One of these differences has already been mentioned—namely, the overall scaling behavior emerges much earlier in the irradiated material with respect to the plastic strain level. This difference is documented in Fig. 11, which presents the strain dependence of the upper scaling limit, $\delta l_{max}$. For each sample, the strain range corresponds to the conditions under which overall scaling behavior is observed. The scaling extent was determined by computing the derivative of the partition functions and identifying the $\delta l$ interval where the derivative fluctuated around a constant level.

More importantly, in light of the discussed contradiction, Fig. 11

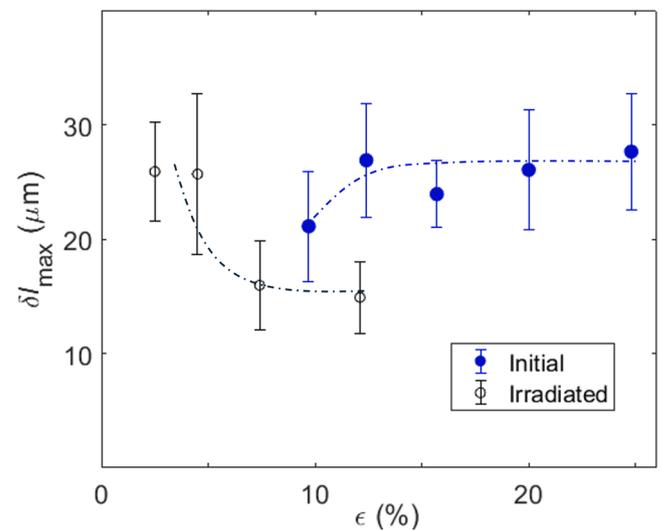

**Fig. 11.** Effect of irradiation on the extent of scale invariance: upper bound of scaling dependencies, $\delta l_{max}$, as a function of strain at which partition functions exhibit global scaling behavior. The dash-dotted lines were drawn arbitrarily as a guide for the eye.

reveals that the $\delta l_{max}(\varepsilon)$ dependencies differ qualitatively between the two cases. For the non-irradiated material, the scaling extent is smallest at 9.7 % strain, increases at the next deformation step (12.4 %), and then remains approximately constant—or perhaps increases slightly—during further deformation. Such a result reflects the slow evolution of the singularity spectra (see Fig. 9(a)). This behavior is consistent with the hypothesis that the characteristic correlation distance is limited primarily by the grain size. It is also worth noting that the relatively large variations in $\delta l_{max}$ and their associated uncertainties provide indirect support for this hypothesis. In addition to purely technical sources of uncertainty (such as the finite size of the KAM images and the limited spatial resolution), variations in the scaling limit may arise from differences in the misorientation evolution among individual grains, which differ in size and orientation relative to the deformation direction.

In contrast, although the irradiated material exhibits a similar scaling extent at small strains, $\delta l_{max}$ decreases with further deformation, reaching considerably smaller values (see vertical lines in Figs. 10(b) and (c)). The saturation length can be compared with the range of distances between slip lines—an internal length that, unlike the grain size, emerges during deformation (see Fig. 7). It can thus be inferred that two competing tendencies exist in the irradiated material: one that extends the scaling limit to larger scales through the development of a hierarchical structure within grains, similar to the non-irradiated sample, and another that limits scaling due to the formation of slip lines, which appear to disrupt this hierarchy. In this context, particular attention is drawn to the high value of $\delta l_{max}$ at 4.5 % strain, although channel nuclei are already visible along grain boundaries at this stage (see Fig. 7(f)). This apparent ambiguity is consistent with the above conjecture. Indeed, the channels discernible in Fig. 7(f) are not yet well developed within most grains and therefore may not constitute strong obstacles to dislocation glide.

## 4. On the nature of mesoscale self-organization of dislocations

Deformation-induced dislocation structures have received significant attention, particularly due to their obvious ordering and a striking degree of spatial organization [42–49,51]. In efforts to describe this visible structure, various features, such as surface roughness or, at finer scales, dislocation structures revealed by TEM characterization have been examined in detail. These analyses have pointed to the self-similarity of the associated patterns across multiple scales. The systematic observation of scaling properties has led to the suggestion that the emergence of fractal features is a universal aspect of the collective dynamics of dislocations [45].

This finding is less self-evident than it may appear. Thus, the dynamics of the dislocations might have been considered uncorrelated, i.e., the system would be treated as a total dislocation population, ignoring their mutual influence. Moreover, uncorrelated dynamics represents the fundamental case in continuum theories of plasticity, where plastic flow is viewed as the result of independent motions of the dislocations through the field of obstacles.

Nevertheless, it is difficult to overlook the highly organized nature of strain-induced dislocation patterns. Understanding the physical origin of the driving force(s) promoting such order is essential for a comprehensive description of the formation of dislocation microstructures. As illustrated in Fig. 1, fractal sets are generated by applying construction rules that ensure self-similarity. This process provides a natural analogy to the emergence of self-organized dislocation structures, where the notion of "construction" can be replaced by "dynamics" and "rules" by "correlations" governed by the interactions between mobile dislocations.

Dislocation systems can involve a large number of rules. These rules may correspond to various phenomena such as long-range and short-range interactions, glide along specific crystallographic planes and directions, climb to other planes, and reactions leading to annihilation, multiplication, or the formation of sessile configurations. Additionally, dislocations move through a stochastic field of pinning obstacles and can be temporarily immobilized. Fundamentally, these factors imply that the rules may be sensitive to both the crystal structure and the defect microstructure. Moreover, the rules themselves may change during deformation, leading to feedback loops. As a result, the fractal properties may vary between materials and evolve with deformation, as has already been reported [41,42,45,46].

Compared with previous studies, this work was intended to interpret an intermediate scale revealed by EBSD analysis of local crystal misorientations induced by a heterogeneous dislocation microstructure. Furthermore, rather than focusing on a specific fractal dimension, the analysis was based on singularity spectra, providing a continuous description of the fractal nature of different subsets within the KAM images as a function of their associated singularity. The observation of finite-width spectra that do not collapse to either monofractal or non-fractal patterns supports the relevance of the chosen approach, as it enables a more comprehensive description of the underlying complexity across all strain levels.

The general conclusion drawn from observations on both non-irradiated and irradiated material is that, on the one hand, dynamical interactions can lead to the emergence of hierarchical dislocation structures, and on the other hand (as indicated by the finite width of the singularity spectra) these structures retain a high degree of heterogeneity even after their formation is complete at sufficiently large strain. At a more detailed level, as illustrated in Fig. 9, the examined evolution of the $f(\alpha)$-dependences clearly reveals specific underlying trends. One of the most direct conclusions from these data is a tendency toward the saturation of changes during deformation. Furthermore, the closeness of the upper scaling limit to the average grain size and its weak strain dependence, as observed in the non-irradiated material (Fig. 11), suggest that changes in dislocation glide orientation at grain boundaries represent a general mechanism limiting the observed scale invariance. These results corroborate and complement earlier findings based on the analysis of selected fractal dimensions and help address some of the uncertainties associated with such analyses [45,46].

An unexpected conclusion emerges from comparing the results of the MF analysis for the irradiated and non-irradiated specimens. At first glance, the respective KAM maps (Fig. 6 and 7) look radically different. The irradiated steel displays conspicuous dislocation channels and coarse slip bands that are absent from the reference material. This visual contrast, however, stems from a single salient feature rather than from a fundamental change in the underlying statistics. The MF response is in fact remarkably similar in the two cases (Fig. 9): the singularity spectra span nearly identical ranges of $\alpha$ and $f$, and their strain-induced evolution follows comparable trajectories, converging toward saturation, as evidenced by the progressive narrowing of the spectra.

This counter-intuitive behavior offers a telling illustration of the additional insight afforded by MF analysis. Moreover, despite the overall similarity of the singularity spectra, the associated scaling behavior reveals meaningful differences. First, although the upper scaling limit for the irradiated sample equals that of the reference material, it decreases progressively with deformation (Fig. 11). Thus, the formation of defect-free channels imposes an extra constraint on scale invariance, whereas the limit corresponds to the average grain size both in the reference material and at low strains in the irradiated specimen. Second, Fig. 11 shows that a hierarchical dislocation structure emerges at considerably lower strains after irradiation, suggesting faster dislocation accumulation outside the defect-free zones. These nuances underscore the high predictive power of MF analysis for the quantitative characterization of KAM maps.

## 5. Concluding remarks

In this study, in situ SEM-EBSD tensile testing combined with MF analysis provided detailed insights into the self-organized nature of deformation-induced dislocation structures in 304 L stainless steel. Applying this methodology to both pristine and neutron-irradiated states enabled a direct comparison of deformation behavior in the same material with intrinsically different microstructures. While plastic deformation of the solution-annealed steel proceeds through the formation of fine slip lines that relatively uniformly fill the grain interiors, irradiation leads to highly heterogeneous plastic flow characterized by the formation of defect-free channels. The comparison of these fundamentally different deformation regimes demonstrates the feasibility of extracting quantitative information on evolving microstructure from SEM-resolution images. At the same time, it advances the understanding of irradiation-induced effects, which, although extensively documented, are still largely assessed on a qualitative basis. Overall, the application of MF analysis to SEM-EBSD data shows strong potential for bridging microscale structural observations with macroscale mechanical behavior.

From a general perspective, extending the analysis from the evaluation of a single fractal dimension to the calculation of singularity spectra allowed to suggest an intrinsically MF nature of the emerging dislocation structures. This behavior implies that their description requires a continuous range of fractal dimensions rather than a single value. Taken together, the results underscore the robustness of the MF analysis in capturing subtle yet essential features of deformation behavior that elude conventional qualitative assessment.

Furthermore, beyond the examination of the MF spectra, important information can be extracted from the analysis of scaling limits for the partition functions (see Eq. (1)) and the evolution of the scaling intervals with strain. In particular, the breakdown of scaling offers insight into competing physical mechanisms—namely, the interplay between dislocation self-organization and the constraints imposed by grain boundaries and slip localization. Notably, the earlier emergence of MF features in the irradiated steel compared with the pristine material, combined with the strain-induced limitation of the scaling intervals due to the formation of defect-free channels, highlights the critical role of irradiation-induced defects in altering deformation pathways and accelerating mesoscale structural evolution.

Finally, it is striking that, despite these pronounced differences between the irradiated and non-irradiated specimens, their singularity spectra exhibit remarkable similarity. This result deserves particular attention because it hints at an underlying commonality in mesoscale dislocation organization. It raises the challenging question of whether common self-organization principles (e.g., universality classes) may govern dislocation dynamics, at least in materials with the same crystal structure. Addressing this question calls for extending the present methodology to a broader range of irradiation conditions, deformation temperatures, and material systems. In this perspective, the quantitative insight provided by the MF analysis may contribute to the development of predictive models for material behavior in extreme environments, such as those encountered in nuclear reactors.

## CRediT authorship contribution statement

**Mikhail Lebyodkin:** Writing – review & editing, Writing – original draft, Visualization, Software, Methodology, Formal analysis, Conceptualization. **Maxim Gussev:** Writing – review & editing, Writing – original draft, Visualization, Resources, Methodology, Investigation, Formal analysis, Data curation, Conceptualization. **Jamieson Brechtl:** Writing – review & editing, Writing – original draft, Visualization, Conceptualization. **Tatiana Lebedkina:** Visualization, Methodology, Formal analysis.

## Declaration of competing interest

The authors declare that they have no known competing financial interests or personal relationships that could have appeared to influence the work reported in this paper.

## Acknowledgments

The authors would like to thank Dr. B. Tanguy (EDF, France) for providing archive 304 L steel. The experimental portion of this research was supported by the Nuclear Science User Facilities (NSUF) Rapid Turnaround Experiment (RTE) award No 19–1698 and by the U.S. Department of Energy, Office of Nuclear Energy, Light Water Reactor Sustainability Program. This manuscript has been authored by UT-Battelle LLC under contract No DE-AC05-00OR22725 with the US Department of Energy (DOE). The US government retains and the publisher, by accepting the article for publication, acknowledges that the US government retains a nonexclusive, paid-up, irrevocable, worldwide license to publish or reproduce the published form of this manuscript, or allow others to do so, for US government purposes. The DOE will provide public access to these results of federally sponsored research in accordance with the DOE Public Access Plan (http://energy.gov/downloads/doe-public-access-plan).